# Digital twins' kinetics of virtual free-radical copolymerization of vinyl monomers with stable radicals. 1. Methyl methacrylate

Elena F. Sheka

Institute of Physical Researches and Technology, Peoples' Friendship University of Russia (RUDN University), 117198 Moscow, Russia;

sheka@icp.ac.ru

**Abstract:** The first experience of virtual free-radical polymerization of a set of vinyl monomers in the framework of the digital twins (DTs) concept (*Polymers* 2023, 15, 2999) is extended to methyl methacrylate, copolymerized with small additives of stable radicals such as fullerene $C_{60}$ and TEMPO. The virtualization of chemical processes is based on the assumption of basic chain reactions that lay the foundation of polymerization. In the current study, a set of elementary reactions, covering the initial stage of the polymerization, is considered. The reactions are the most suitable for quantum chemical treatment. The calculations, covering about 30 DTs, were carried out using a semi-empirical version of the unrestricted two-determinant Hartree-Fock approximation. The main energy and spin-density parameters of the ground state of DTs are determined. The barrier profiles of the decomposition of three DTs were calculated, which led the foundation for determining the activation energy of the studied processes. The decisive role of spins in the formation of transition states is confirmed. The two stable radicals behave quite differently. The action of fullerene $C_{60}$ concerns mainly the capturing of the initiating free radicals, anchoring them to the molecule body and thus slowing the main polarization process. In contrast, *TEMPO* effectively capture monomer radicals manifesting itself as a 'killer' of the polymerization thus providing the appearance of the induction period in practice. The obtained virtual kinetic data are in a full consent with experimental reality.

## 1. Introduction

Free-radical polymerization (FRP) of vinyl monomers is the most studied chemical process and the introduction of weak additions of complementary stable radicals into their reactors has always been under the constant attention of researchers [1]. Once focused on optimizing the technology for obtaining the final polymer, the researchers, having not found a noticeable effect of these radicals on the final product, came to a consensus about their weak practical significance. The situation changed drastically when fullerene $C_{60}$ was introduced into the polymerization reactor as an additional stable radical (see monographic review [2] and references therein). The effect turned out to be so strong that authors of Ref. [3] had to speak about observing the free-radical copolymerization (FRCP) of the studied monomer (xylylene) with $C_{60}$. The feature was supported with other researchers and very soon, FRCP turned into a broad area of new polymer science. The issues has been studied by several groups and for various vinyl monomers (xylylene [3], styrene [4 – 12], methyl methacrylate [7,13, 14], 4-vinylbenzoic acid [15], maleic anhydride [16], etc.). Studies of this first wave of activity revealed two main features that accompany the involving of fullerene in the FRCP. The first concerns the effect of the fullerene-provoked inhibition of the reaction at its initial stage while the second is related to branched-star liked structures of polymer molecules anchored on fullerene. Studies of the second wave of the activity [17-21] were mainly aimed at elucidating the details of the inhibition-fullerene kinetics, with respect to which a wide range of opinions was observed. The vision of the final stage of the reaction, concerning star formation [22], has been generally accepted unchanged until now.

The discrepancy in understanding the inhibitory role of $C_{60}$ fullerene in FPCR is naturally associated with the fundamental basis of this process, which is figuratively well described with a 'pineapple-on-plantation' picture [23] caused by a multivariable intermolecular interaction between molecular substances that fill a reaction solution (RS). The result of each of the potential pair interactions may prevent polymer chains growth from being a winner of the strong competition. The latter is controlled by the kinetics of the reactions occurring and determines the winner with the fastest kinetics, all other things being equal. 'Ceteris paribus' includes such main reaction parameters as solvent, concentration of initial reagents, temperature, RS design and so forth. On the other hand, distinctive characteristics of the reactions such as the types of initiating and stable radicals as well as monomers are carriers of the required information to be necessary for the determination of the thermodynamic and kinetic parameters of the considered reactions.

In spite of a large variation of the results obtained, there is a possibility to reveal certain trends concerning the inhibitory effects of stable radicals on FPR. Let's consider these trends using the example of reactions carried out with the maximum approximation to the 'ceteris paribus' format. Fundamental studies [20, 21] involved two different neat monomers (methyl methacrylate (MMA) and styrene (St)), two initiating free radicals $AIBN^{\bullet}$ and $BP^{\bullet}$ produced in the course of the thermal decomposition of either 2,2'-azobisisobutyronitrile (*AIBN*) or benzoyl peroxide (*BP*), respectively, as well as two complementary stable radicals $C_{60}$ and *TEMPO*. The obtained results concerning the kinetics of the initial stage of the FRP of the mentioned monomers have allowed to reveal three main trends. The first concerns the FRCP of monomer with $C_{60}$, the conversion rate of which reduces by several time in the fullerene presence with respect to that one of the neat monomer. This trend is characteristic for the ($C_{60}$ + MMA) FRCP. The second trend is associated with the presence of the induction period (IP) at the initial stage of the studied FRCP on the background of practically unchanged conversion rate of the neat monomer. The trend is observed for the (*TEMPO*+MMA), (*TEMPO*+St) as well as ($C_{60}$ + St) FRCPs. The third trend concerns the similarity and difference in the reactions of the two neat monomers on the presence of stable radicals $C_{60}$ and *TEMPO*. The main goal of the current study is to replace these verbal characteristics with digital kinetic parameters that allow to numerically describing the observed empirical patterns.

The revealed trends provide rich food for thought, on the one hand, and clearly outline the field of activity of virtual polymerization. Obviously, the virtual confirmation of the selected trends concerns the foundations of the FRCP mechanism. The implementation of a virtual game, involving all the discussed trends and previously not realistic, today is possible within the framework of a new modeling paradigm - the Digital Twins (DTs) one [23-25]. This approach makes it possible to consider the totality of the presented trends under conditions 'other things being equal' and to reveal those features of the stable radicals that determine their voluminous role in the FRP of vinyl monomers. The successful implementation of the game paradigm is determined by how clearly the playing field is configured and the game characters are defined.

## 2. Playing field and characters of the Digital Twins' virtual FRCP game

The configuration of the playing field in this work is based on the chain-reaction concept laying the foundation of the complex polymerization process that thus presenting it as a well-traced sequence of superpositional elementary reactions [26,27]. From a theoretical viewpoint, such a vision of polymerization is the most favorable for using the quantum-chemical (QC) techniques for its virtual consideration, reducing it to the consideration of individual elementary reactions [28,29]. The theory of elementary reactions and their QC consideration has been going on for many decades [30-34], and the only complaint about the certain limitations of these studies can be the fact that highly welcome QC calculations of sets of one-type reactions did not become widespread. The main novelty of the DT concept concerns just this key point since a large massive of elementary reactions, which is the playing field in the current study, allows to clearly distinguish one-type reactions, performed under the same conditions, followed by a comparative analysis of their results, accompanied by the establishment of reliable trends. The status of 'the same conditions' implies the application to the same QC consideration, absolute temperature zero, and vacuum medium.

**Table 1**. Nomination of elementary reactions and/or digital twins related to the initial stage of the free-radical copolymerization of vinyl monomers with stable radicals

| Reaction mark | Reaction equation [1] | Reaction rate constant | Reaction type |
|---|---|---|---|
| (1) | $R^\bullet + M \to RM^\bullet$ | $k_i$ | generation of monomer-radicals |
| (2) | $RM^\bullet + (n-1)M \to RM_n^\bullet$ | $k_p$ | generation of oligomer-radicals, polymer chain growth |
| (3a) | $F + M \to FM$ | $k_{2m}^F$ | two-dentant grafting of monomer on $C_{60}$ |
| (3b) | $F + M \to FM^\bullet$ | $k_{1m}^F$ | one-dentant stable radical grafting of monomer, generation of monomer-radical |
| (4) | $FM^\bullet + (n-1)M \to FM_n^\bullet$ | $k_p^F$ | generation of oligomer-radical anchored to $C_{60}$, polymer chain growth |
| (5) | $S + M \to SM^\bullet \equiv SM$ | $k_{1m}^S$ | one-dentant coupling with monomer |
| (6) | $F + RM^\bullet \to FRM$ | $k_{rm}^F$ | monomer-radical grafting on $C_{60}$ |
| (7) | $S + RM^\bullet \to SRM$ | $k_{rm}^S$ | monomer-radical capturing with stable radical |
| (8) | $F + R^\bullet \to FR$ | $k_R^F$ | free radical grafting on $C_{60}$ |
| (9) | $S + R^\bullet \to SR$ | $k_R^F$ | free radical capturing with stable radical |
| (10) | $F + S \to FS$ | $k_S^F$ | stable radical grafting on $C_{60}$ |
| (11) | $R^\bullet + FM^\bullet \to RFM$ | $k_{FM}^R$ | monomer-radical $FM^\bullet$ capturing with free radical |
| (12) | $S + FM^\bullet \to SFM$ | $k_{FM}^S$ | monomer-radical $FM^\bullet$ capturing with radical S |

[1] $M, R, F, S$ mark a vinyl monomer, initiating free radicals (either $AIBN^\bullet$ or $BP^\bullet$), stable radicals (fullerene $C_{60}$, and *TEMPO*), respectively. Superscript black spot distinguishes radical participants of the relevant reactions.

The characters of the virtual game under discussion involve a large number of DTs, which are participants in diverse elementary reactions. The latter are schematically presented in Table 1 in terms of the relevant DTs. The reaction list does not present all of the potential elementary events, say, reactions of chain transfer and other are skipped, while being quite complete in view of the empirical trends under study.

The first common characteristic of the reactions listed in Table 1 is their radical character. However, they are therewith distinctly divided into two groups that cover association reactions (1), (2), (5) and (7), uniting either free or stable radical with monomer, and the remaining grafting reactions that in the case of the stable-radical $C_{60}$ are reactions of the fullerene derivatization of different kinds. Over thirty years ago, the latter products were called fullerenyls [35]. The past decades since then, this name has taken root [36-39] and we will use it in the future. Products of reactions (1), (2), (3b), and (4) are free radicals while all others are either stable species or fullerenyl stable radicals in the case of *TEMPO* and $C_{60}$, respectively.

Reaction (1) $RM^\bullet$ is the cornerstone of the entire polymerization process, determining its feasibility as such. By selecting the most successful participants in this reaction empirically, the researchers opted for monotarget free radicals such as $AIBN^\bullet$ and $BP^\bullet$, while the stable radical *TEMPO* was found unsuitable for this role. As for $C_{60}$, the first wave of polymer researchers, who introduced fullerene into polymerization and were confident in its radical nature, made repeated attempts to detect reaction (3b), accompanied by the growth of a polymer chain attached to the fullerene (reaction (4)) [3-16, 40]. The fate of reaction (3b) depends on the detailed configuration of the intermolecular junction between $C_{60}$ and a monomer. The latter is configured with two $sp^2$C-C bonds, one belonging to fullerene and the other presenting a vinyl group of monomer. Accordingly, the junction can be either two-dentantly or one-dentantly configured. If the first configuration causes the formation of a [2x2] cycloaddition monoadduct fullerenyl $FM$ similar to the patterned $C_{60}$, the second results in the formation of fullerene-grafted monomer radical $FM^\bullet$ similar to $RM^\bullet$. Accordingly, reaction (2) describes the polymer chain growth initiated with a free radical while reaction (4) describes the monomer polymerization, once grafted on fullerene. Although the existence of a reaction (3b) was suspected in a number of cases, a confident conclusion was not made and this reaction as well as reaction (4) were classified as unlikely.

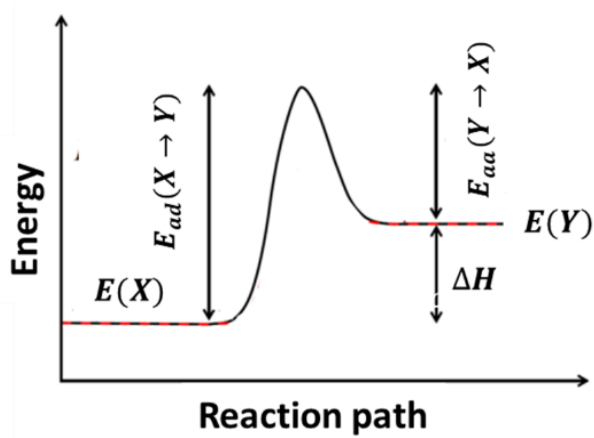

**Figure 1.** Universal energy graph of the pairwise intermolecular interaction in the system of many possibilities.

The elementary reaction concept of FRP/FRCP provides one more common thing. This concerns a standard energy graph. Actually, each elementary reaction proceeds between a pair of reagents, the energy of intermolecular interaction between which follows a typical graph when the components of the pair get closer in the chosen direction, taken as the reaction coordinate (see Figure 1). As seen in the figure, the graph includes the total energy of the equilibrium reactant community $Y$, $E(Y)$, the energy of the product of the $Y$ pair interaction, $E(X)$, and the energy of the transition state of the molecular complex under consideration, $E_{TS}(X \leftrightarrow Y)$. As for kinetics of the reaction, its standard description concerns the rate constant, $k(T)$, which is expressed through the Arrhenius equation (Eq. 1) [31-34]

$$k(T) = Ae^{\left(\frac{-E_a}{kT}\right)}. \tag{1}$$

Here $A$ is a complex frequency factor, while $E_a$ presents the activation energy, which is either the energy of the $X$ product decomposition, $E_{ad}$, or the $Y$ pair association, $E_{aa}$. There is also one more important energetic parameter – reaction enthalpy, $\Delta H$, or coupling energy, $E_{cpl}=E(X)-E(Y)$.

The main difficulty in evaluation of the $k(T)$ value is highly complicated nature of the frequency factor $A$. Its determination concerns basic problems of rotational-vibrational dynamics of polyatomic molecules, such as great number of both vibrational and rotational degrees of freedom as well as their anharmonicity. However, for one-type elementary reactions, A factor is expected to change weakly [28, 31-34], so that activation energy becomes governing. Its value can be determined by building barrier profiles of either association or decomposition of molecular pairs $Y$ and $X$, respectively.

## 3. Digital twins concept. Digital twins of reagents under discussion

A general algorithm of the DT concept can be presented schematically as [24,25]

*Digital twins → Virtual device → IT product*.

Here, DTs are molecular models under study, virtual device is a carrier of a selected software, IT product covers a large set of computational results related to the DTs under different actions in the light of the soft explored. The quality of the product highly depends on how broadly and deeply the designed DTs cover the knowledge concerning the object under consideration and how adequate is the virtual device to peculiarities of this object. The concept is completely free from statistical and random errors that accompany real experiments, and is favored to any modification.  IT product in the current study presents structural, thermodynamic and kinetic parameters that in terms of the universal energy graph accompany any of the elementary reactions discussed in the previous section.

Each virtual experiment concerns a virtual reaction solution (VRS), the constituents of which form a pool of the studied DTs. The latters of the current study present vinyl monomer MMA; free radical $AIBN^\bullet$ and $BP^\bullet$; two stable radicals - *TEMPO* and fullerene $C_{60}$; and a number of final products, among which various monoadduct fullerenyls, designed for each elementary reaction particularly. Virtual device in the current study is the CLUSTER-Z1 software [41,42] implementing AM1 version of the semi-empirical unrestricted two-determinant Hartree-Fock (UHF) approach [43]. The program showed itself highly efficient concerning open-shell electronic systems such as fullerenes [44,45], graphene molecules [46], and stable radicals [47,48]. A detailed discussion concerning the choice of proper softwares for virtual FRP of vinyl monomers is presented elsewhere [23].  Digital twins of fullerenyls were designed basing on spin chemistry of fullerene $C_{60}$ [44,45], a short sketch of which is given in the Supporting Material.

## 4. Virtual free-radical copolymerization of methyl methacrylate with stable radicals

### 4.1. Fundamental grounds

A set of cross-elementary reactions in the form of a matrix, the boundaries of which are mapped with their participants, turned out to be the most convenient for the simultaneous presentation and discussion of a large number of possibilities and results. In a certain form, matrix Table 2 is a visualization of a 'pineapple-on-the-plantation' [23] in the realities of the FRP of methyl methacrylate and its copolymerization with C$_{60}$ fullerene and *TEMPO*. The corresponding 'matrix elements' are evidently divided into four groups marked with different colors. Yellow elements present elementary reactions and their final products that govern initial stage of the MMA FRP. Elements in light blue describe reactions and final products related to the FRCP of MMA with *TEMPO*. Faint-pink elements do the same with respect to fullerene C$_{60}$. Individual light-gray element is related to the interaction of *TEMPO* with C$_{60}$. The first third of the table lists final products of paired elementary reactions, following the accepted designation of both reactions and their products in Table 1. Duplicate matrix elements are omitted. The matrix as a whole describes the case when the FRP of monomer is mainly provided with free radical-initiator $AIBN^\bullet$, while the FRCP of the monomer occurs because of introduction of *TEMPO* and C$_{60}$ into the VRS. As seen from this part of the table, reactions and/or products can be divided in two groups. Members of the first group are rooted on monomer (M) and concern the generation of different monomer-based radicals listed in the first row. Those are dimer radical $R^AM_2^\bullet$, ensuring the growth of the polymer chain of the monomer M, two monomer radicals $R^AM^\bullet$ and $SM$, which determine the beginning of the chain growth due to the initiator radicals $AIBN^\bullet$ ($R^{A\bullet}$) and *TEMPO* ($S^\bullet$), as well as the monomer radical $FM^\bullet$, which determines potential initiation of the generation and growth of the polymer chain anchored to the fullerene.

**Table 2.** Elementary reactions and DTs of their final products supplemented with virtual thermodynamic and kinetic descriptors related to the FRCP of methyl methacrylate with stable radicals *TEMPO* and C$_{60}$

| | $R^AM^{\bullet 1)}$ | $AIBN^\bullet$ $R^{A\bullet 1)}$ | TEMPO $S^\bullet$ | C$_{60}$ $F$ | |
|---|---|---|---|---|---|
| | | | | 2-dentant | 1-dentant |
| $M$ | $R^AM_2^{\bullet\bullet}$ | $R^AM^\bullet$ | $SM$ | $FM$ | $FM^\bullet$ |
| $R^AM^\bullet$ | | | $SR^AM$ | $FR^AM$ | |
| $R^{A\bullet}$ | - | - | $SR^A$ | $FR^A$ | |
| $R^{T\bullet}$ | - | - | - | $FS$ | |
| **Coupling energies, $E_{cpl}$, kcal/mol** | | | | | |
| $M$ | -15.62 (2) [1)] <br> -6.95 (3) <br> -18.98 (4) <br> -11.55 (5) | -14.56 (C) [2)] <br> 1.68 (CH$_2$) <br> $N_{DA}$= 0.96 e | not formed | -19.10 | 29.75 <br> $N_{DA}$= 0.96 e |
| $R^AM^\bullet$ | | | --2.79 | -16.11 | |
| $R^{A\bullet}$ | - | - | 5.001 | -20.14 | |
| $R^{T\bullet}$ | - | - | - | 0.98 | |
| **Activation energies, $E_a$, kcal/mol** | | | | | |
| $M$ | 10.46 (2) | 12.28 (1) | not formed | - | $E_{ad} \gg E_{ac}$ |
| $R^AM^\bullet$ | - | - | 17.63 | 11.58 | |
| $R^{A\bullet}$ | - | - | $E_{ad} > E_{ac}$ | 9.40 | |
| $R^{T\bullet}$ | - | - | - | not formed | |

[1)] Digits in brackets mark the number of monomers in the oligomer chain.
[2)] Atomic compositions in brackets mark the target location on the vinyl bond of the monomer-radical.

All other matrix elements are related to the FRCP of MMA with the two stable radicals. Thus, the reactions and/or products of the second row express a threat for a complete forbiddance of the polymer chain growth caused by capturing the leading monomer radical $R^A M^\bullet$. The first reaction $(R^A M)_2$ concerns its dimerization. Reactions $R^A R^A M$ and $SR^A M$ describe the loss of the radical ability of the monomer radical $R^A M^\bullet$ because of attracting free radicals $R^{A\bullet}$ and $S^\bullet$, respectively. Reaction $FR^A M$ corresponds the case when $C_{60}$ is acting as a killer of the FRP. Three last members describe the interaction of the main free radical $R^{A\bullet}$ with other two, which evidently leads to their additional consumption and, consequently, to the FRCP slowing.

The second third of the table presents the thermodynamic deriptors of the reactions described above in terms of their energies, $\Delta H$, or coupling energies, $E_{cpl}$, of their final products. The last third does the same but for kinetic descriptors in terms of the activation energies of the corresponding reactions, $E_{aa} = E_a$. The bold data are obtained by constructing barrier profiles of the decomposition of the final products of the relevant elementary reactions. Curve 1 in Figure 2 describes the barrier profile of the $SR^A M$ species decomposition, thus revealing the activation energy 'killing' the kinetics of the first step of the MMA polymerization propagation. Curve 2 plots the main parameter that is responsible for this propagation, while curve 3 exhibits the activity of fullerene $C_{60}$ to capture free radical $R^{A\bullet}$. Other data in the top-left matrix cells are evaluated when using Evans-Polanyi-Semenov (EPS) relations [23,32,49] that linearly couple $E_{cpl}$ and $E_a$ of one-type elementary reactions providing the polymerization of vinyl monomers [23]. As shown, these relations in the case of reactions generating monomer-radicals and dimer-radicals occurred reliable. Actually, the EPS-based $E_a$ value for the MMA dimer-radical of the current study (top data in the first cell of the bottom part of Table 2) is well consistent with that one followed from the barrier profile shown in Figure 2.

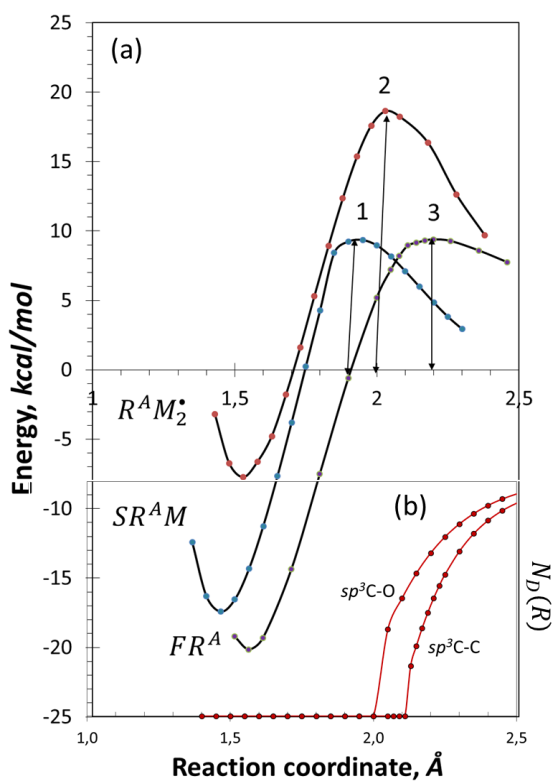

**Figure 2.** (a) Barrier profiles of the decomposition of $SR^A M$ (1), dimer radical $R^A M_2^\bullet$ (2) and $FR^A$ (3) products (see text). (b). $N_D(R)$ graphs, related to the elongation of the $sp^3$C-C bond of ethane and $sp^3$C-O bond of ethylene glycol [51]. UHF AM1 calculations.

In the previous study [23], where the detailed view of energy graphs as barrier profiles was given for the first time, a spin origin of transition states was established. The latter was evidenced by the fact that position of the energy graphs maxima coincides with $R_{crit}^{C-C}$ of 2.11±0.1 Å that determines the maximum length of the $sp^3$C-C bond, above which the bond becomes radicalized thus revealing the start of its breaking [50,51]. This bond played the role of the reaction coordinate in all the cases studied earlier [23]. As seen in Figure 2, the same in related to the $R^A M_2^\bullet$ and $FR^A$ cases, the barrier-profile maxima of which are located at 2.01 and 2.08 Å. Oppositely to the case, the $sp^3$C-O bond is central in the case of intermolecular junctions in the $SR^A M$ species. Expectedly, the maximum positions of its barrier profile should differ from those provide with breaking of the $sp^3$C-C bond. Actually, these positions constitute 1.97 Å, which well correlates with $R_{crit}^{C-O}$ of 2.05 Å relating to the dissociation of the $sp^3$C-O bond of ethylene glycol presented in the bottom of Figure 2. Evidently, $R_{crit}^{C-O}$, as well as $R_{crit}^{C-C}$, deviates in different atomic surrounding, which is proved by the values presented above.

### 4.2. Virtual and real kinetics of initial stage of the FRP of methyl methacrylate and its FRCP with *TEMPO* and C$_{60}$

The numerical descriptors of virtual thermodynamics and kinetics of the FRP of methyl methacrylate and its FRCP with *TEMPO* and C$_{60}$, collected in this study are presented in Table 2. The equilibrium DTs structures, associated with the matrix elements of the table are divided into two groups related to the FRCP of MMA with C$_{60}$ only (Figure 3) and corresponding to the case when both stable radicals are involved of the polymerization process (Figure 4). Graphical inserts in the figures present experimental summary data on the kinetics of the relevant processes.

Molecular structures in Figure 3a present virtual FRP of MMA, starting with creation of monomer radical $R^A M^\bullet$ and proceeding with sequential generation oligomer radicals $R^A M_{n+1}^\bullet$ with $n$ from 1 to 3. This group of DTs is described with yellow data of Table 2. Evidently, in general, MMA FRP is fully similar to that typical to all other single-group vinyl monomers [23] by both complicated structure configuring when growing $n$ and energy parameters $E_{cpl}$ and $E_a$ listed in Table 2, while differing in details. The MMA vinyl group forms $sp^2$C-CH$_2$ bond, the carbon atom of methylene group of which willingly generates a stable intermolecular $sp^3$C-C bond with the target carbon of the $AIBN^\bullet$ in contrast to a sole carbon atom, coupling of which with $R^{A\bullet}$ target is thermodynamically not profitable. The intermolecular junction between free radical and monomer is provided with the mentioned $sp^3$C-C bond, while a sole carbon atom of the vinyl becomes a carrier of the radicalized atom, servicing as a new target for the next attacking action. A detailed description of all the feature of vinyl monomer FRP can be found in Ref. 23.

When fullerene $F$ joins the company of $M$ and $R^{A\bullet}$, intermolecular interaction stimulates its contact with the two partners first, result of which is exhibited in Figure 4b. Evidently, the $M$ and $F$ contact is provided with tight cooperation of $sp^2$C-C bonds of the MMA vinyl group and one of the C$_{60}$ such bonds. As discussed earlier, this two-bond contact can be either [2x2] cycloaddition two-dentant or one-dentant. The first is resulted in the formation of a stable fullerene monoadduct fullerenyl $FM$, the reaction thermodynamic of which is quite favorable. Radical properties of the species are governed by the pool of unpaired electrons $N_D$, modified by the monomer anchoring and distributed over all the cage atoms [45]. Concerning the polymerization, this product presents just a side effect causing a negligible consumption of the monomer due to small concentration of C$_{60}$. However, when the interbond contact between monomer and fullerene is one-dentant, the MMA vinyl bond reacts on the fullerene presence similarly to that of a free radical such as $R^{A\bullet}$, generating a typical free radical of the alone carbon atom of the group thus forming a fullerene based monomer radical $FM^\bullet$. The radicalization of the latter constitutes 0.96 e, so that this radical is highly similar to such as $R^A M^\bullet$ and, ideally, can lead to the propagation of the monomer polymerization. Evidently, the decisive word belongs to the energy parameters. As seen in Table 2, $E_{cpl}$ of the monomer radical $FM^\bullet$ is positive indicating that its formation is thermodynamically not profitable. Additionally, $E_{cpl}$ large value means that the corresponding energy graph is characterized with $E_{ad} \gg E_{aa}$, making the formation of the radical kinetically impossible.

The interaction of C$_{60}$ with free radical $R^{A\bullet}$ occurs much simpler and results in the formation of a standard fullerene derivative - monofullerenyl $FR^A$, radical properties of which is governed with a modified pool $N_D$, presented in Figure S1b. Therefore, capturing of free radicals with fullerene in the current VRS is thermodynamically quite favorable. In contrast, a similar action of the fullerene with respect to monomer radical $R^A M^\bullet$ (reaction $FR^A M$) was unsuccessful, since none of the numerous attempts to link this radical with C$_{60}$ was positive so that none of stable structural formations $FR^A M$ was obtained.

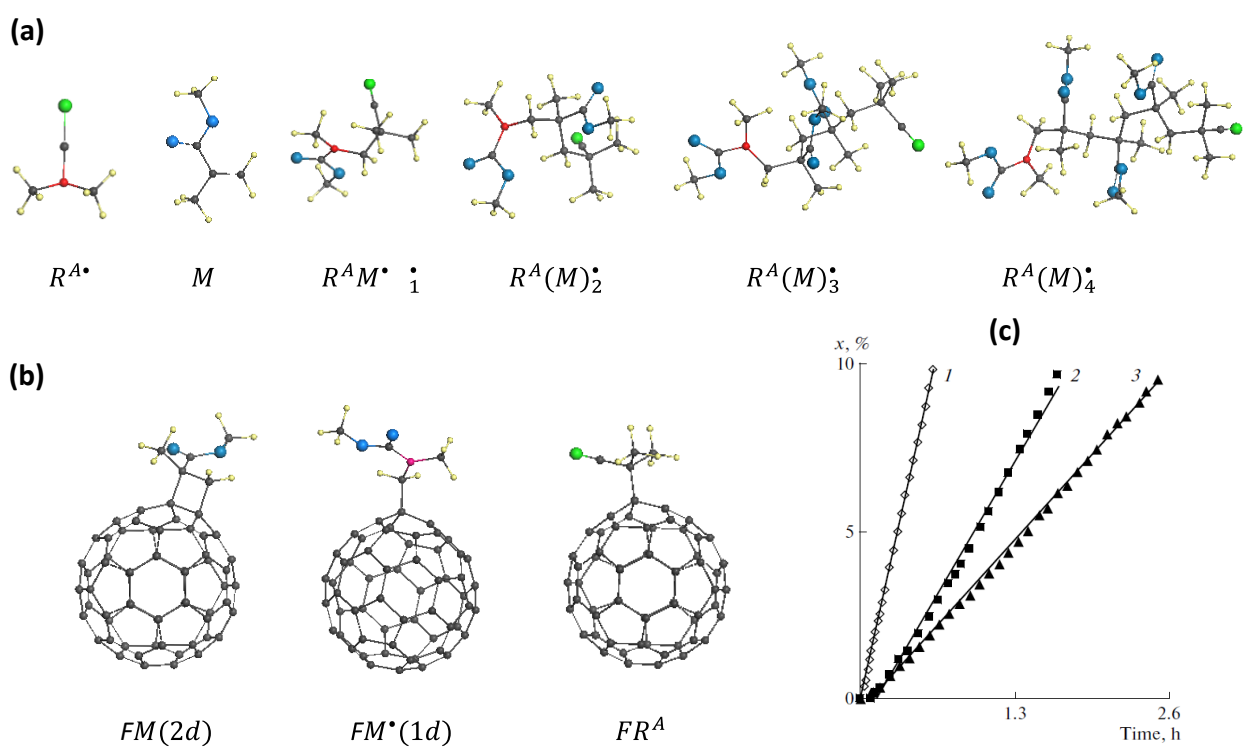

**Figure 3**. Equilibrium structures of digital twins of FRCP MMA with fullerene C$_{60}$. (a) Oligomer radicals $R(M)^\bullet_{n+1}$ for $n$ from 1 till 3. (b) Monofullerenyls $FM$, $FM^\bullet$, and $FR^A$. Small yellow, gray and red balls mark hydrogen, common and target carbon atoms, respectively. Larger green and blue balls depict nitrogen and oxygen atoms. UHF AM1 calculations. (c) Kinetics of MMA polymerization initiated with 2.0 × 10$^{-2}$ mol/L AIBN at 333 K in the presence of different fullerene C$_{60}$ concentrations of (*1*) 0, (*2*) 1.0 × 10$^{-3}$, and (*3*) 2.0 × 10$^{-3}$ mol/L. Digitalized data from [20].

Turning to a discussion of the kinetics of the considered elementary reactions in the $E_a$ -part of Table 2, we should pay attention to the data presented in yellow and faint pink. As seen from the table, the former process, related to the FRP of MMA, is governed with activation energy filling the range from 11 to 24 kcal/mol. The values are typical for all the studied vinyl monomers [23] and are kinetically quite favorable for the experimental implementation of their FRP to be successful. Actually, curve 1 in Figure 3c represents the empirical percentage monomer conversion $x(t)$ of the MMA monomer array in due course of its FRP initiated with $AIBN^\bullet$. The corresponding elementary reactions include the initiation of the monomer radical and the successive growth of the polymer chain. Rates of the corresponding reactions, $k_i$ and $k_p$, respectively, can be considered as reference.

Coming back to Table 2, we could not ignore the fact that in the current reaction solution occurs one more process, namely, capturing the acting free radicals with fullerene $C_{60}$, activation energy of which is less of any former ones promoting FRP of MMA. The equilibrium structure of the relevant final product $FR^A$ is shown in Figure 3b. This kinetically favorable event may strongly decrease the current concentration of free radicals, thus slowing the main FRP. It is this slowing that is clearly seen in Figure 3c, when a small fraction of $C_{60}$, equal to 1/100 of the monomer mass, is added to the reactor. As can be seen from the figure, the linear growth of the reference FRP conversion is preserved, but its rate is three times less. Obviously, increase in the $C_{60}$ concentration lifts the rate of free radicals removal from the MMA polymerization, which is confirmed experimentally by a new about twofold decrease in the monomer conversion rate with a twofold increase in the $C_{60}$ concentration (cf. curves 2 and 3 in Figure 3c). The $R^{A\bullet}$ anchoring can be not only single, but also numerous. Therefore, the role of $C_{60}$ in its FRCP with MMA is to adsorb free radicals $R^{A\bullet}$ in due course of an extended $R^{A\bullet}$-polyderivatization of $C_{60}$. The kinetical predominance of this elementary reaction is exhibited as retaining a straight-linearly dependent conversion of monomer while remarkably lowering its rate.

The complication of the reference FRP of MMA by addition of another stable radical $S^\bullet$ instead of $C_{60}$ leads to a completely different behavior of the MMA polymerization. According to the data, listed in Table 2 and marked with light blue, radical $S^\bullet$ does not initiate the formation of monomer radical $SM^\bullet$ (Figure 4a), capable of polymerizing the monomer independently from $R^{A\bullet}$. The relevant species $SM$ is inactive. In contrast to $C_{60}$, the *TEMPO*'s ability to capture free radicals $R^{A\bullet}$ (see Figure 4b) is not thermodynamically favorable as well and is kinetically unreal due to $E_{ad} > E_{aa}$. The only meaningful elementary reaction, concerning the triad, consisting of monomer and two radicals $R^{A\bullet}$ and $S^\bullet$, concerns reaction $SR^A M$- the removing of the monomer radical $R^A M^\bullet$ from the further polymerization process by its adsorbing with radical $S^\bullet$. The final product of this reaction is shown in Figure 4c and reveals a strong coupling of radical $S^\bullet$ with monomer radical $R^A M^\bullet$. Thus, 'killing' the monomer radicals $R^A M^\bullet$ is the main role of the *TEMPO* in the FRCP of MMA with *TEMPO*. Evidently, the killing prevents from the propagation of the MMA polymer chain thus terminating the monomer polymerization, which provides the appearance of the IP on the conversion vs time plotting. Evidently, the IP duration is determined with the radical $S^\bullet$ concentration since the FRP of MMM cannot start until all this radical mass is consumed. Thereafter, a standard FRP of MMA will proceed, which is revealed with the same conversion rate of the reference and post-performed processes (cf. curves 1 and 1' in Figure 4e).

The joint addition of *TEMPO* and $C_{60}$ into the VRS is expected to be a superposition of effects presented with curves 1 and 2 of plotting in Figure 3c and with curves 1 and 1' plotted in Figure 4e if there is no reaction between the two radicals. As seen in Table 2, a potential reaction $FS$ does not occur (see the reaction final product in Figure 4d) so that the mentioned superposition is highly probable. As seen in Figure 4e, it is really observed and pairs of curves 1-1 and 2-2' just duplicate each other, while differing with the conversion rates, the same for each pair, caused by the fullerene presence. Taking as a whole, the experimental picture, covering FRP of MMA as well as FRCPs of MMA with *TEMPO* and $C_{60}$, reliably shows that elementary reactions of the FRP of MMA themselves and in the presence of additional stable radicals such as *TEMPO* and $C_{60}$ are independent and, thus, superpositional.

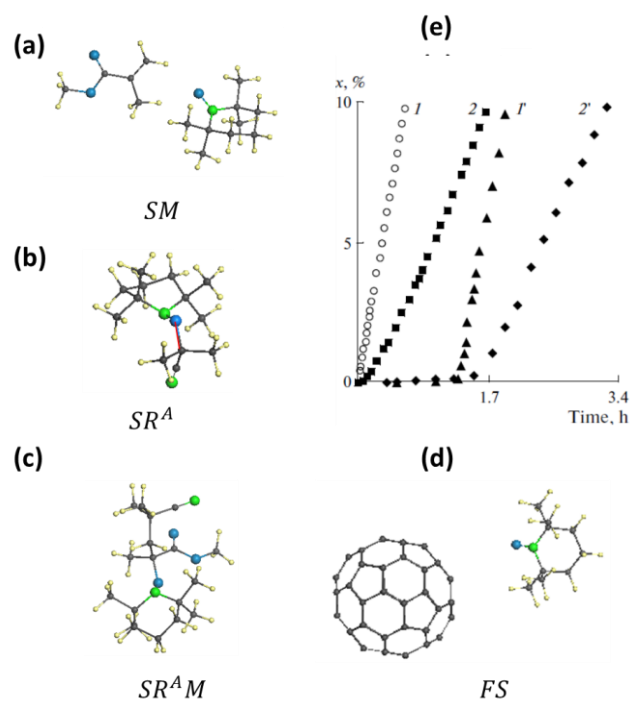

**Figure 4**. Free-radical copolymerization of MMA with $C_{60}$ and *TEMPO*. Equilibrium structures of digital twins related to elementary reactions $SM$ (a), $SR^A$ (b), $SR^A M$ (c), and $FS$ (d) (see the reactions nomination in Tables 1 and 2). Small yellow and gray balls mark hydrogen and common carbon atoms, respectively. Larger green and blue balls depict nitrogen and oxygen atoms. UHF AM1 calculations. (e) Conversion of the MMA in the course of polymerization initiated with $2.0 \times 10^{-2}$ mol/L AIBN at 333 K. (*1*) In the absence of TEMPO, (*1'*) in the presence of $1.0 \times 10^{-3}$ mol/L TEMPO, (*2*) in the presence of $1.0 \times 10^{-3}$ mol/L of fullerene $C_{60}$, but the absence of TEMPO and (*2'*) in the presence of $1.0 \times 10^{-3}$ mol/L of both fullerene $C_{60}$ and TEMPO. Digitalized data from [20].

## 5. Conclusion

The concept of digital twins, which has successfully demonstrated itself in the virtualization of free-radical polymerization of vinyl monomers [23], is used in this work to determine the mechanism of the influence of additional stable radicals on the above process. The success of the implementation of the concept is determined by the reliability of the basic concept considering polymerization as a chain reaction consisting of a set of independent elementary reactions occurred superpositionally. In this view, polymerization is a product of a won kinetic competition in the environment of a set of elementary reactions. The quantum chemical approximation used suggested the activation energy as a kinetic descriptor for labeling final reaction products.

In this work, the family of vinyl monomers is represented by methyl methacrylate, which undergoes rapid polymerization when initiated by the free radical $AIBN^\bullet$. Additives of stable radicals were provided with *TEMPO* and fullerene $C_{60}$. The latter stimulate their copolymerization with the monomer, which has a significant impact on the course of the main FRP reaction. The FRP of the monomer and its FRCP with stable radicals were divided into a network of elementary reactions, each of which was numerically analyzed in terms of a standard energy graph of reaction theory.

The results obtained are as follows.

1. A virtual examination of the full set of elementary reactions gives a complete picture of the mechanism of FRP and FRCP in the coordinates of the chemical reagents used.

2. MMA FRP, virtualized for the first time in this work, is allowed and proceeds in a similar manner with respect to other vinyl monomers.

3. Despite the large selection of potential interactions and reactions in a virtual reaction solution, fullerene additives are most kinetically favorable only for the capture and adsorption of $AIBN^\bullet$ free radicals. This capture affects the number of free radicals and leads to a decrease in the conversion rate of the main monomer in the presence of fullerene. Moreover, an increase in fullerene concentration is accompanied by a cymbate further decrease in conversion, which is reliably observed experimentally.

4. Unlike fullerene, for *TEMPO* the most kinetically favorable reaction is the capture of a monomer radical. This entrapment prevents propagation of monomer polymerization, which leads to the occurrence of an induction period on the temporary conversion plot of the monomer. In full accordance with this prediction, the experimental dependences of the monomer conversion in the case of its FRCP with *TEMPO* show an extended induction period.

5. An explanation of the reasons that determine the experimentally observed multiple decrease in the conversion rate of MMA in the presence of $C_{60}$ additives and the appearance of an extended induction period in the MMA FRCP in the presence of *TEMPO* has been explained for the first time.

6. In the initial period of polymerization, the single-target nature of *TEMPO* in contrast to the multi-target nature of fullerene $C_{60}$ do not have a fundamental effect on the polymerization process. However, if in the case of *TEMPO* the radical is fully worked out in the initial period, then in the case of $C_{60}$, the fullerenyls formed in the initial period continue to participate in the further formation of the final polymer product, providing their multitarget carbon structure for anchoring growing polymer chains.

**Digital twins' kinetics of virtual free-radical copolymerization of vinyl monomers with stable radicals. 1. Methyl methacrylate**

Elena F. Sheka

Institute of Physical Researches and Technology, Peoples' Friendship University of Russia (RUDN University), 117198 Moscow, Russia;

sheka@icp.ac.ru

**Supporting Material**

Digital twins of fullerenyls were designed basing on spin chemistry (SCh) of fullerene $C_{60}$ [1-5]. According to the latter, the molecule is a stable radical carrying unpaired of the total number $N_D$ = 9.6 $e$. These electrons are distributed over the carbon atoms of the molecule in accordance with the partial fractions $N_{DA}$ on each atom, represented by the histogram in Figure S1a. $N_{DA}$ is curve), and free valence $V_{free}$ (curve with dots). Different colors in the insert distinguish six atomic groups shown by the $Z \rightarrow A$ graph. (b). $sp^2$C-C bond distribution of the fullerene C$_{60}$ (histogram) and fullerenyl $FR^A$ (black dotted curve). Insert presents equilibrium structure of the species, see Table 2. Red curve plots $N_{DA}$ of the fullerenyl. UHF AM1 calculations.

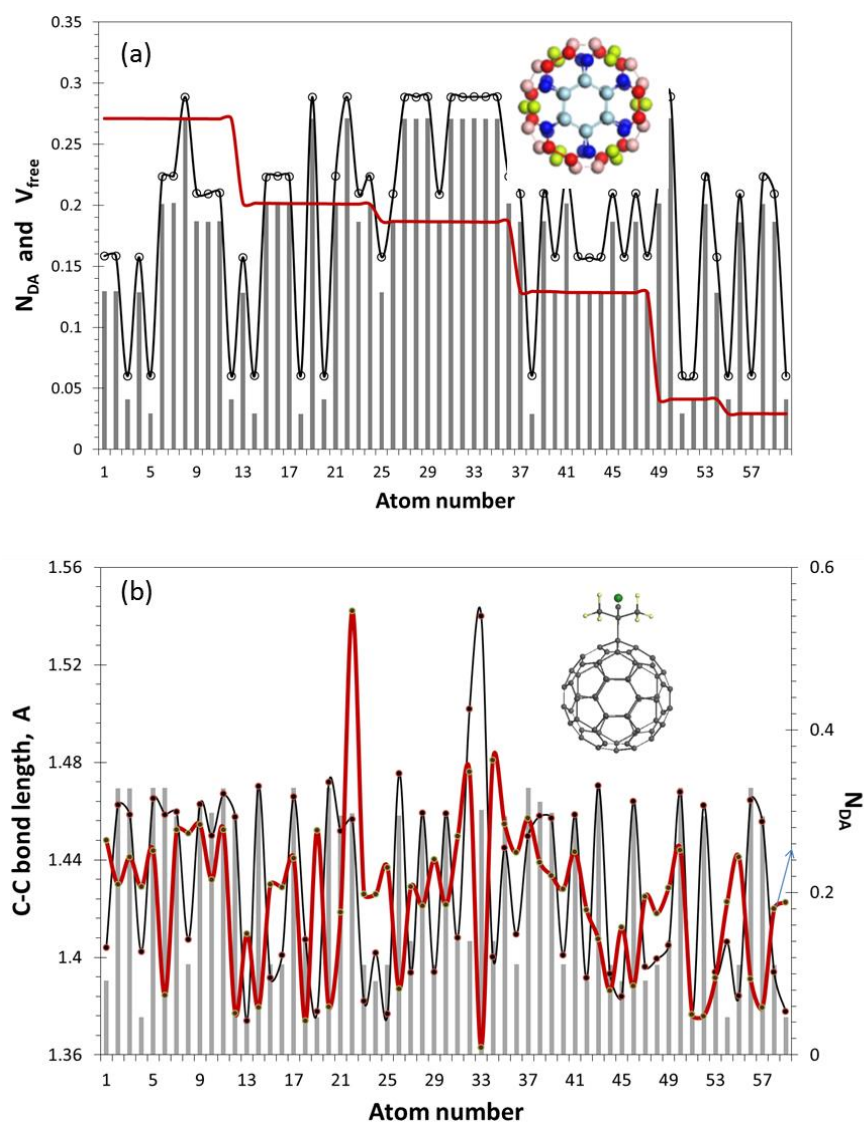

**Figure S1**. (a). Atomic chemical susceptibility $N_{DA}$ of fullerene $C_{60}$, distributed over the molecule atoms according to either their numeration in the output file (histogram) or in the $Z \rightarrow A$ manner (stepped red

a quantitative indicator of the atomic chemical susceptibility (ACS) of atom A, which is in full agreement with the corresponding free valence of the atom, $V_A^{free}$ (see curve with dots in the figure). Arranged in the *max→min* format (stepped red curve), they reveal five groups of atoms, each consisting of 12 atoms and characterized by a constant $N_{DA}$ value within the group. These groups of atoms are shown in different colors in the structural insert of the figure. The light gray atoms, which form two central hexagons, represent six identical $sp^2$C-C pairs, the first to enter any type of reaction involving C$_{60}$. One of the main concepts of the fullerene' SCh is the controlling role of $N_{DA}$ over the process of stepwise polyderivatization of the fullerene molecule [1-5]. Tunable after each completed step of attaching the corresponding addend, the $N_{DA}$ distribution reveals the maximum value that determines the target atom for the next attack. Several examples of such computational polyderivatization of C$_{60}$ fullerene, confirmed by experimental data, are given in the monograph [4]. In what follows, the matter will be about various C$_{60}$ monoderivatives. In all the cases, atom 33 was selected as a target one. As seen in Figure S1b, anchoring the addend to this atom causes a drastic elongation of one covalent bond (black curve) transferring it from $sp^2$-type of 1.423 Å to $sp^3$ one of 1.534 Å. Because of the closed structure of the neat molecule, this elongation is accompanied with a reconstruction of all the covalent bonds, which, remaining of $sp^2$-type, change their lengths. The latter is resulted in the reconstruction of the $N_{DA}$ distribution of the neat C$_{60}$ (red curve), thus revealing the highest value of 0.55 *e* at atom 22 that is a partner to atom 33 of a broken $sp^2$C-C bond. Atom 22 becomes the best target for the next anchoring and preserves this role for small addends such as individual atoms, OH, COOH and other groups [4]. Molecular addends may not be added to this atom because of sterical constraints. The target atoms, suitable for the addition, should be looked through the next $N_{DA}$ top-list atoms, anchoring to which is free from the sterical constraints. When the selection is over and the necessary computations are performed, newly reconstructed $N_{DA}$ distribution is analyzed to make choice the next potential targets among the top-list cage atoms suitable for the next addition without sterical constraints. The next-step derivatization proceeds similarly until the total pool of unpaired electrons $N_D$ is consumed [1-5].